# Efficient and robust approximate nearest neighbor search using Hierarchical Navigable Small World graphs

Yu. A. Malkov, D. A. Yashunin

**Abstract** — We present a new approach for the approximate K-nearest neighbor search based on navigable small world graphs with controllable hierarchy (Hierarchical NSW, HNSW). The proposed solution is fully graph-based, without any need for additional search structures, which are typically used at the coarse search stage of the most proximity graph techniques. Hierarchical NSW incrementally builds a multi-layer structure consisting from hierarchical set of proximity graphs (layers) for nested subsets of the stored elements. The maximum layer in which an element is present is selected randomly with an exponentially decaying probability distribution. This allows producing graphs similar to the previously studied Navigable Small World (NSW) structures while additionally having the links separated by their characteristic distance scales. Starting search from the upper layer together with utilizing the scale separation boosts the performance compared to NSW and allows a logarithmic complexity scaling. Additional employment of a heuristic for selecting proximity graph neighbors significantly increases performance at high recall and in case of highly clustered data. Performance evaluation has demonstrated that the proposed general metric space search index is able to strongly outperform previous opensource state-of-the-art vector-only approaches. Similarity of the algorithm to the skip list structure allows straightforward balanced distributed implementation.

**Index Terms** — Graph and tree search strategies, Artificial Intelligence, Information Search and Retrieval, Information Storage and Retrieval, Information Technology and Systems, Search process, Graphs and networks, Data Structures, Nearest neighbor search, Big data, Approximate search, Similarity search

——————————— ◆ ———————————

## 1 INTRODUCTION

Constantly growing amount of the available information resources has led to high demand in scalable and efficient similarity search data structures. One of the generally used approaches for information search is the K-Nearest Neighbor Search (K-NNS). The K-NNS assumes you have a defined distance function between the data elements and aims at finding the $K$ elements from the dataset which minimize the distance to a given query. Such algorithms are used in many applications, such as non-parametric machine learning algorithms, image features matching in large scale databases [1] and semantic document retrieval [2]. A naïve approach to K-NNS is to compute the distances between the query and every element in the dataset and select the elements with minimal distance. Unfortunately, the complexity of the naïve approach scales linearly with the number of stored elements making it infeasible for large-scale datasets. This has led to a high interest in development of fast and scalable K-NNS algorithms.

Exact solutions for K-NNS [3-5] may offer a substantial search speedup only in case of relatively low dimensional data due to "curse of dimensionality". To overcome this problem a concept of Approximate Nearest Neighbors Search (K-ANNS) was proposed, which relaxes the condition of the exact search by allowing a small number of errors. The quality of an inexact search (the recall) is defined as the ratio between the number of found true nearest neighbors and $K$. The most popular K-ANNS solutions are based on approximated versions of tree algorithms [6, 7], locality-sensitive hashing (LSH) [8, 9] and product quantization (PQ) [10-17]. Proximity graph K-ANNS algorithms [10, 18-26] have recently gained popularity offering a better performance on high dimensional datasets. However, the power-law scaling of the proximity graph routing causes extreme performance degradation in case of low dimensional or clustered data.

In this paper we propose the Hierarchical Navigable Small World (Hierarchical NSW, HNSW), a new fully graph based incremental K-ANNS structure, which can offer a much better logarithmic complexity scaling. The main contributions are: explicit selection of the graph's enter-point node, separation of links by different scales and use of an advanced heuristic to select the neighbors. Alternatively, Hierarchical NSW algorithm can be seen as an extension of the probabilistic skip list structure [27] with proximity graphs instead of the linked lists. Performance evaluation has demonstrated that the proposed general metric space method is able to strongly outperform previous opensource state-of-the-art approaches suitable only for vector spaces.

————————————————

- *Y. Malkov is with the Federal state budgetary institution of science Institute of Applied Physics of the Russian Academy of Sciences, 46 Ul'yanov Street, 603950 Nizhny Novgorod, Russia. E-mail: yurymalkov@mail.ru.*
- *D. Yashunin. Addres: 31-33 ul. Krasnovezdnaya, 603104 Nizhny Novgorod, Russia. E-mail: yashuninda@yandex.ru*

## 2 RELATED WORKS

### 2.1 Proximity graph techniques

In the vast majority of studied graph algorithms searching takes a form of greedy routing in k-Nearest Neighbor (k-NN) graphs [10, 18-26]. For a given proximity graph, we start the search at some enter point (it can be random or supplied by a separate algorithm) and iteratively traverse the graph. At each step of the traversal the algorithm examines the distances from a query to the neighbors of a current base node and then selects as the next base node the adjacent node that minimizes the distance, while constantly keeping track of the best discovered neighbors. The search is terminated when some stopping condition is met (e.g. the number of distance calculations). Links to the closest neighbors in a k-NN graph serve as a simple approximation of the Delaunay graph [25, 26] (a graph which guranties that the result of a basic greedy graph traversal is always the nearest neighbor). Unfortunately, Delaunay graph cannot be efficiently constructed without prior information about the structure of a space [4], but its approximation by the nearest neighbors can be done by using only distances between the stored elements. It was shown that proximity graph approaches with such approximation perform competitive to other k-ANNS thechniques, such as kd-trees or LSH [18-26].

The main drawbacks of the k-NN graph approaches are: 1) the power law scaling of the number of steps with the dataset size during the routing process [28, 29]; 2) a possible loss of global connectivity which leads to poor search results on clusetered data. To overcome these problems many hybrid approaches have been proposed that use auxiliary algorithms applicable only for vector data (such as kd-trees [18, 19] and product quantization [10]) to find better candidates for the enter nodes by doing a coarse search.

In [25, 26, 30] authors proposed a proximity graph K-ANNS algorithm called Navigable Small World (NSW, also known as Metricized Small World, MSW), which utilized *navigable* graphs, i.e. graphs with logarithmic or polylogarithmic scaling of the number of hops during the greedy traversal with the respect of the network size [31, 32]. The NSW graph is constructed via consecutive insertion of elements in random order by bidirectionally connecting them to the $M$ closest neighbors from the previously inserted elements. The $M$ closest neighbors are found using the structure's search procedure (a variant of a greedy search from multiple random enter nodes). Links to the closest neighbors of the elements inserted in the beginning of the construction later become bridges between the network hubs that keep the overall graph connectivity and allow the logarithmic scaling of the number of hops during greedy routing.

Construction phase of the NSW structure can be efficiently parallelized without global synchronization and without mesuarable effect on accuracy [26], being a good choice for distributed search systems. The NSW approach delivered the state-of-the-art performance on some datasets [33, 34], however, due to the overall polylogarithmic complexity scaling, the algorithm was still prone to severe performance degradation on low dimensional datasets (on which NSW could lose to tree-based algorithms by several orders of magnitude [34]).

### 2.2 Navigable small world models

Networks with logarithmic or polylogarithmic scaling of the greedy graph routing are known as the navigable small world networks [31, 32]. Such networks are an important topic of complex network theory aiming at understanding of underlying mechanisms of real-life networks formation in order to apply them for applications of scalable routing [32, 35, 36] and distributed similarity search [25, 26, 30, 37-40].

The first works to consider spatial models of navigable networks were done by J. Kleinberg [31, 41] as social network models for the famous Milgram experiment [42]. Kleinberg studied a variant of random Watts-Strogatz networks [43], using a regular lattice graph in d-dimensional vector space together with augmentation of long-range links following a specific long link length distribution $r^{-\alpha}$. For $\alpha$=d the number of hops to get to the target by greedy routing scales polylogarithmically (instead of a power law for any other value of $\alpha$). This idea has inspired development of many K-NNS and K-ANNS algorithms based on the navigation effect [37-40]. But even though the Kleinberg's navigability criterion in principle can be extended for more general spaces, in order to build such a navigable network one has to know the data distribution beforehand. In addition, greedy routing in Kleinberg's graphs suffers from polylogarithmic complexity scalability at best.

Another well-known class of navigable networks are the scale-free models [32, 35, 36], which can reproduce several features of real-life networks and advertised for routing applications [35]. However, networks produced by such models have even worse power law complexity scaling of the greedy search [44] and, just like the Kleinberg's model, scale-free models require global knowledge of the data distribution, making them unusable for search applications.

The above-described NSW algorithm uses a simpler, previously unknown model of navigable networks, allowing decentralized graph construction and suitable for data in arbitrary spaces. It was suggested [44] that the NSW network formation mechanism may be responsible for navigability of large-scale biological neural networks (presence of which is disputable): similar models were able to describe growth of small brain networks, while the model predicts several high-level features observed in large scale neural networks. However, the NSW model also suffers from the polylogarithmic search complexity of the routing process.

## 3 MOTIVATION

The ways of improving the NSW search complexity can be identified through the analysis of the routing process, which was studied in detail in [32, 44]. The routing can be divided into two phases: "zoom-out" and "zoom-in" [32]. The greedy algorithm starts in the "zoom-out" phase from a low degree node and traverses the graph simultaneously increasing the node's degree until the characteristic radius of the node links length reaches the scale of the distance to the query. Before the latter happens, the average degree of a node can stay relatively small, which leads to an increased probability of being stuck in a distant false local minimum.

One can avoid the described problem in NSW by starting the search from a node with the maximum degree (good candidates are the first nodes inserted in the NSW structure [44]), directly going to the "zoom-in" phase of the search. Tests show that setting hubs as starting points substantially increases probability of successful routing in the structure and provides significantly better performance at low dimensional data. However, it still has only a polylogarithmic complexity scalability of a single greedy search at best, and performs worse on high dimensional data compared to Hierarchical NSW.

The reason for the polylogarithmic complexity scaling of a single greedy search in NSW is that the overall number of distance computations is roughly proportional to a product of the average number of greedy algorithm hops by the average degree of the nodes on the greedy path. The average number of hops scales logarithmically [26, 44], while the average degree of the nodes on the greedy path also scales logarithmically due to the facts that: 1) the greedy search tends to go through the same hubs as the network grows [32, 44]; 2) the average number of hub connections grows logarithmically with an increase of the network size. Thus we get an overall polylogarithmic dependence of the resulting complexity.

The idea of Hierarchical NSW algorithm is to separate the links according to their length scale into different layers and then search in a multilayer graph. In this case we can evaluate only a needed fixed portion of the connections for each element independently of the networks size, thus allowing a logarithmic scalability. In such structure the search starts from the upper layer which has only the longest links (the "zoom-in" phase). The algorithm greedily traverses through the elements from the upper layer until a local minimum is reached (see Fig. 1 for illustration). After that, the search switches to the lower layer (which has shorter links), restarts from the element which was the local minimum in the previous layer and the process repeats. The maximum number of connections per element in all layers can be made constant, thus allowing a logarithmic complexity scaling of routing in a navigable small world network.

One way to form such a layered structure is to explicitly set links with different length scales by introducing layers. For every element we select an integer level $l$ which defines the maximum layer for which the element belongs to. For all elements in a layer a proximity graph (i.e. graph containing only "short" links that approximate Delaunay graph) is built incrementally. If we set an exponentially decaying probability of $l$ (i.e. following a geometric distribution) we get a logarithmic scaling of the expected number of layers in the structure. The search procedure is an iterative greedy search starting from the top layer and finishing at the zero layer.

In case we merge connections from all layers, the structure becomes similar to the NSW graph (in this case the $l$ can be put in correspondence to the node degree in NSW). In contrast to NSW, Hierarchical NSW construction algorithm does not require the elements to be shuffled before the insertion - the stochasticity is achieved by using level randomization, thus allowing truly incremental indexing even in case of temporarily altering data distribution (though changing the order of the insertion slightly alters the performace due to only partially determenistic construction procedure).

The Hierarchical NSW idea is also very similar to a well-known 1D probabilistic skip list structure [27] and can be described using its terms. The major difference to skip list is that we generalize the structure by replacing the linked list with proximity graphs. The Hierarchical

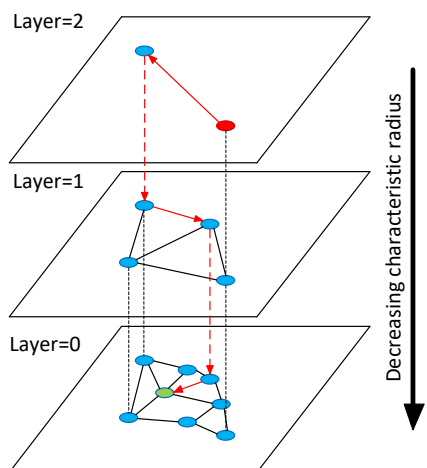

Fig. 1. Illustration of the Hierarchical NSW idea. The search starts from an element from the top layer (shown red). Red arrows show direction of the greedy algorithm from the entry point to the query (shown green).

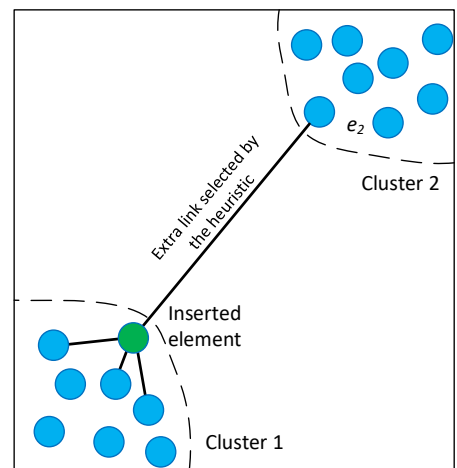

Fig. 2. Illustration of the heuristic used to select the graph neighbors for two isolated clusters. A new element is inserted on the boundary of Cluster 1. All of the closest neighbors of the element belong to the Cluster 1, thus missing the edges of Delaunay graph between the clusters. The heuristic, however, selects element $e_2$ from Cluster 2, thus, maintaining the global connectivity in case the inserted element is the closest to $e_2$ compared to any other element from Cluster 1.

NSW approach thus can utilize the same methods for making the distributed approximate search/overlay structures [45].

For the selection of the proximity graph connections during the element insertion we utilize a heuristic that takes into account the distances between the candidate elements to create diverse connections (a similar algorithm was utilized in the spatial approximation tree [4] to select the tree children) instead of just selecting the closest neighbors. The heuristic examines the candidates starting from the nearest (with respect to the inserted element) and creates a connection to a candidate only if it is closer to the base (inserted) element compared to any of the already connected candidates (see Section 4 for the details).

When the number of candidates is large enough the heuristic allows getting the exact relative neighborhood graph [46] as a subgraph, a minimal subgraph of the Delaunay graph deducible by using only the distances between the nodes. The relative neighborhood graph allows easily keeping the global connected component, even in case of highly clustered data (see Fig. 2 for illustration). Note that the heuristic creates extra edges compared to the exact relative neighborhood graphs, allowing controlling the number of the connections which is important for search performance. For the case of 1D data the heuristic allows getting the exact Delaunay subgraph (which in this case coincides with the relative neighborhood graph) by using only information about the distances between the elements, thus making a direct transition from Hierarchical NSW to the 1D probabilistic skip list algorithm.

Base variant of the Hierarchical NSW proximity graphs was also used in ref. [18] (called 'sparse neighborhood graphs') for proximity graph searching. Similar heuristic was also a focus of the FANNG algorithm [47] (published shortly after the first versions of the current manuscript were posted online) with a slightly different interpretation, based on the sparse neighborhood graph's property of the exact routing [18].

## 4 ALGORITHM DESCRIPTION

Network construction algorithm (alg. 1) is organized via consecutive insertions of the stored elements into the graph structure. For every inserted element an integer maximum layer $l$ is randomly selected with an exponentially decaying probability distribution (normalized by the $m_L$ parameter, see line 4 in alg. 1).

The first phase of the insertion process starts from the top layer by greedily traversing the graph in order to find the $ef$ closest neighbors to the inserted element $q$ in the layer. After that, the algorithm continues the search from the next layer using the found closest neighbors from the previous layer as enter points, and the process repeats. Closest neighbors at each layer are found by a variant of the greedy search algorithm described in alg. 2, which is an updated version of the algorithm from [26]. To obtain the approximate $ef$ nearest neighbors in some layer $l_c$, a dynamic list $W$ of $ef$ closest found elements (initially filled with enter points) is kept during the search. The list is updated at each step by evaluating the neighborhood of the closest previously non-evaluated element in the list until the neighborhood of every element from the list is evaluated. Compared to limiting the number of distance calculations, Hierarchical NSW stop condition has an advantage - it allows discarding candidates for evalution that are further from the query than the furthest element in the list, thus avoiding bloating of search structures. As in NSW, the list is emulated via two priority queues for better performance. The distinctions from NSW (along with some queue optimizations) are: 1) the enter point is a fixed parameter; 2) instead of changing the number of multi-searches, the quality of the search is controlled by a different parameter $ef$ (which was set to $K$ in NSW [26]).

**Algorithm 1**
INSERT($hnsw$, $q$, $M$, $M_{max}$, $efConstruction$, $m_L$)
**Input**: multilayer graph $hnsw$, new element $q$, number of established connections $M$, maximum number of connections for each element per layer $M_{max}$, size of the dynamic candidate list $efConstruction$, normalization factor for level generation $m_L$
**Output**: update $hnsw$ inserting element $q$
1  $W \leftarrow \emptyset$  // list for the currently found nearest elements
2  $ep \leftarrow$ get enter point for $hnsw$
3  $L \leftarrow$ level of $ep$   // top layer for $hnsw$
4  $l \leftarrow \lfloor -\ln(unif(0..1)) \cdot m_L \rfloor$  // new element's level
5  **for** $l_c \leftarrow L \ldots l+1$
6      $W \leftarrow$ SEARCH-LAYER($q$, $ep$, $ef$=1, $l_c$)
7      $ep \leftarrow$ get the nearest element from $W$ to $q$
8  **for** $l_c \leftarrow \min(L, l) \ldots 0$
9      $W \leftarrow$ SEARCH-LAYER($q$, $ep$, $efConstruction$, $l_c$)
10     $neighbors \leftarrow$ SELECT-NEIGHBORS($q$, $W$, $M$, $l_c$)  // alg. 3 or alg. 4
11     add bidirectionall connections from $neighbors$ to $q$ at layer $l_c$
12     **for** each $e \in neighbors$  // shrink connections if needed
13         $eConn \leftarrow neighbourhood(e)$ at layer $l_c$
14         **if** $|eConn| > M_{max}$ // shrink connections of $e$
                             // if $l_c = 0$ then $M_{max} = M_{max0}$
15             $eNewConn \leftarrow$ SELECT-NEIGHBORS($e$, $eConn$, $M_{max}$, $l_c$)
                             // alg. 3 or alg. 4
16             set $neighbourhood(e)$ at layer $l_c$ to $eNewConn$
17     $ep \leftarrow W$
18  **if** $l > L$
19      set enter point for $hnsw$ to $q$

**Algorithm 2**
SEARCH-LAYER($q$, $ep$, $ef$, $l_c$)
**Input**: query element $q$, enter points $ep$, number of nearest to $q$ elements to return $ef$, layer number $l_c$
**Output**: $ef$ closest neighbors to $q$
1  $v \leftarrow ep$   // set of visited elements
2  $C \leftarrow ep$  // set of candidates
3  $W \leftarrow ep$  // dynamic list of found nearest neighbors
4  **while** $|C| > 0$
5      $c \leftarrow$ extract nearest element from $C$ to $q$
6      $f \leftarrow$ get furthest element from $W$ to $q$
7      **if** $distance(c, q) > distance(f, q)$
8          **break**   // all elements in $W$ are evaluated
9      **for** each $e \in neighbourhood(c)$ at layer $l_c$  // update $C$ and $W$
10         **if** $e \notin v$
11             $v \leftarrow v \cup e$
12             $f \leftarrow$ get furthest element from $W$ to $q$
13             **if** $distance(e, q) < distance(f, q)$ or $|W| < ef$
14                 $C \leftarrow C \cup e$
15                 $W \leftarrow W \cup e$
16                 **if** $|W| > ef$
17                     remove furthest element from $W$ to $q$
18  **return** $W$

**Algorithm 3**
SELECT-NEIGHBORS-SIMPLE(*q*, *C*, *M*)
**Input**: base element *q*, candidate elements *C*, number of neighbors to return *M*
**Output**: *M* nearest elements to *q*
**return** *M* nearest elements from *C* to *q*

**Algorithm 5**
K-NN-SEARCH(*hnsw*, *q*, *K*, *ef*)
**Input**: multilayer graph *hnsw*, query element *q*, number of nearest neighbors to return *K*, size of the dynamic candidate list *ef*
**Output**: *K* nearest elements to *q*
1  $W \leftarrow \emptyset$  // set for the current nearest elements
2  $ep \leftarrow$ get enter point for *hnsw*
3  $L \leftarrow$ level of *ep*  // top layer for *hnsw*
4  **for** $l_c \leftarrow L \dots 1$
5    $W \leftarrow$ SEARCH-LAYER(*q*, *ep*, *ef*=1, $l_c$)
6    $ep \leftarrow$ get nearest element from *W* to *q*
7  $W \leftarrow$ SEARCH-LAYER(*q*, *ep*, *ef*, $l_c$ =0)
8  **return** *K* nearest elements from *W* to *q*

During the first phase of the search the *ef* parameter is set to 1 (simple greedy search) to avoid introduction of additional parameters.

When the search reaches the layer that is equal or less than *l*, the second phase of the construction algorithm is initiated. The second phase differs in two points: 1) the *ef* parameter is increased from 1 to *efConstruction* in order to control the recall of the greedy search procedure; 2) the found closest neighbors on each layer are also used as candidates for the connections of the inserted element.

Two methods for the selection of *M* neighbors from the candidates were considered: simple connection to the closest elements (alg. 3) and the heuristic that accounts for the distances between the candidate elements to create connections in diverse directions (alg. 4), described in the Section 3. The heuristic has two additional parameters: *extendCandidates* (set to **false** by default) which extends the candidate set and useful only for extremely clustered data, and *keepPrunedConnections* which allows getting fixed number of connection per element. The maximum number of connections that an element can have per layer is defined by the parameter $M_{max}$ for every layer higher than zero (a special parameter $M_{max0}$ is used for the ground layer separately). If a node is already full at the moment of making of a new connection, then its extended connection list gets shrunk by the same algorithm that used for the neighbors selection (algs. 3 or 4).

The insertion procedure terminates when the connections of the inserted elements are established on the zero layer.

The K-ANNS search algorithm used in Hierarchical NSW is presented in alg. 5. It is roughly equivalent to the insertion algorithm for an item with layer *l*=0. The difference is that the closest neighbors found at the ground layer which are used as candidates for the connections are now returned as the search result. The quality of the search is controlled by the *ef* parameter (corresponding to *efConstruction* in the construction algorithm).

**Algorithm 4**
SELECT-NEIGHBORS-HEURISTIC(*q*, *C*, *M*, $l_c$, *extendCandidates*, *keepPrunedConnections*)
**Input**: base element *q*, candidate elements *C*, number of neighbors to return *M*, layer number $l_c$, flag indicating whether or not to extend candidate list *extendCandidates*, flag indicating whether or not to add discarded elements *keepPrunedConnections*
**Output**: *M* elements selected by the heuristic
1  $R \leftarrow \emptyset$
2  $W \leftarrow C$  // working queue for the candidates
3  **if** *extendCandidates*   // extend candidates by their neighbors
4    **for** each $e \in C$
5      **for** each $e_{adj} \in neighbourhood(e)$ at layer $l_c$
6        **if** $e_{adj} \notin W$
7          $W \leftarrow W \cup e_{adj}$
8  $W_d \leftarrow \emptyset$  // queue for the discarded candidates
9  **while** $|W| > 0$ **and** $|R| < M$
10   $e \leftarrow$ extract nearest element from *W* to *q*
11   **if** *e* is closer to *q* compared to any element from *R*
12     $R \leftarrow R \cup e$
13   **else**
14     $W_d \leftarrow W_d \cup e$
15 **if** *keepPrunedConnections*  // add some of the discarded
                                    // connections from $W_d$
16   **while** $|W_d| > 0$ **and** $|R| < M$
17     $R \leftarrow R \cup$ extract nearest element from $W_d$ to *q*
18 **return** *R*

## 4.1 Influence of the construction parameters

Algorithm construction parameters $m_L$ and $M_{max0}$ are responsible for maintaining the small world navigability in the constructed graphs. Setting $m_L$ to zero (this corresponds to a single layer in the graph) and $M_{max0}$ to *M* leads to production of directed k-NN graphs with a power-law search complexity well studied before [21, 29] (assuming using the alg. 3 for neighbor selection). Setting $m_L$ to zero and $M_{max0}$ to infinity leads to production of NSW graphs with polylogarithmic complexity [25, 26]. Finally, setting $m_L$ to some non-zero value leads to emergence of controllable hierarchy graphs which allow logarithmic search complexity by introduction of layers (see the Section 3).

To achieve the optimum performance advantage of the controllable hierarchy, the overlap between neighbors on different layers (i.e. percent of element neighbors that are also belong to other layers) has to be small. In order to decrease the overlap we need to decrease the $m_L$. However, at the same time, decreasing $m_L$ leads to an increase of average hop number during a greedy search on each layer, which negatively affects the performance. This leads to existence of the optimal value for the $m_L$ parameter.

A simple choice for the optimal $m_L$ is $1/\ln(M)$, this corresponds to the skip list parameter $p=1/M$ with an average single element overlap between the layers. Simulations done on an Intel Core i7 5930K CPU show that the proposed selection of $m_L$ is a reasonable choice (see Fig. 3 for data on 10M random d=4 vectors). In addition, the plot demonstrates a massive speedup on low dimensional data when increasing the $m_L$ from zero and the effect of using the heuristic for selection of the graph connections. It is hard to expect the same behavior for high dimensional data since in this case the k-NN graph already has very short greedy algorithm paths [28]. Surprisingly, increasing the $m_L$ from zero leads to a measurable increase in speed on very high dimensional data (100k dense random d=1024 vectors, see plot in Fig. 4), and does not introduce any penalty for the Hierarchical NSW approach. For real data such as SIFT vectors [1] (which have complex mixed structure), the performance improvement by increasing the $m_L$ is higher, but less prominent at current settings compared to improvement from the heuristic (see Fig. 5 for 1-NN search performance on 5 million 128-dimensional SIFT vectors from the learning set of BIG-ANN [13]).

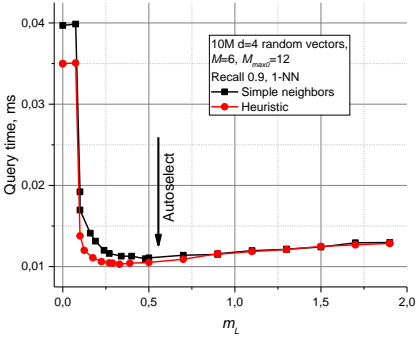

Fig. 3. Plots for query time vs $m_L$ parameter for 10M random vectors with d=4. The autoselected value $1/\ln(M)$ for $m_L$ is shown by an arrow.

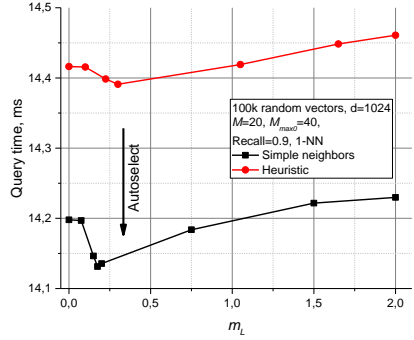

Fig. 4. Plots for query time vs $m_L$ parameter for 100k random vectors with d=1024. The autoselected value $1/\ln(M)$ for $m_L$ is shown by an arrow.

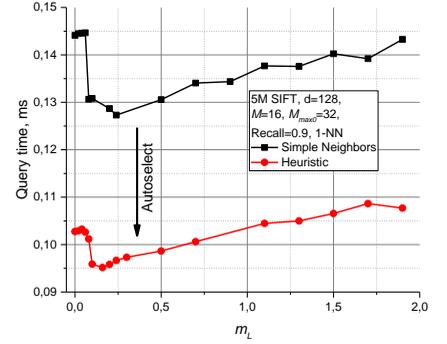

Fig. 5. Plots for query time vs $m_L$ parameter for 5M SIFT learn dataset. The autoselected value $1/\ln(M)$ for $m_L$ is shown by an arrow.

Selection of the $M_{max0}$ (the maximum number of connections that an element can have in the zero layer) also has a strong influence on the search performance, especially in case of high quality (high recall) search. Simulations show that setting $M_{max0}$ to $M$ (this corresponds to k-NN graphs on each layer if the neighbors selection heuristic is not used) leads to a very strong performance penalty at high recall. Simulations also suggest that $2 \cdot M$ is a good choice for $M_{max0}$; setting the parameter higher leads to performance degradation and excessive memory usage. In Fig. 6 there are presented results of search performance for the 5M SIFT learn dataset depending on the $M_{max0}$ parameter (done on an Intel Core i5 2400 CPU). The suggested value gives performance close to optimal at different recalls.

In all of the considered cases, use of the heuristic for proximity graph neighbors selection (alg. 4) leads to a higher or similar search performance compared to the naïve connection to the nearest neighbors (alg. 3). The effect is the most prominent for low dimensional data, at high recall for mid-dimensional data and for the case of highly clustered data (ideologically discontinuity can be regarded as a local low dimensional feature), see the comparison in Fig. 7 (Core i5 2400 CPU). When using the closest neighbors as connections for the proximity graph, the Hierarchical NSW algorithm fails to achieve a high recall for clustered data because the search stucks at the clusters boundaries. Contrary, when the heuristic is used (together with candidates' extension, line 3 in Alg. 4), clustering leads to even higher performance. For uniform and very high dimensional data there is a little difference between the neighbors selecting methods (see Fig. 4), possibly due to the fact that in this case almost all of the nearest neighbors are selected by the heuristic.

The only meaningful construction parameter left for the user is $M$. A reasonable range of $M$ is from 5 to 48. Simulations show that smaller $M$ generally produces better results for lower recalls and/or lower dimensional data, while bigger $M$ is better for high recall and/or high dimensional data (see Fig. 8 for illustration, Core i5 2400 CPU). The parameter also defines the memory consumption of the algorithm (which is proportional to $M$), so it should be selected with care.

Selection of the *efConstruction* parameter is straightforward. As it was suggested in [26] it has to be large enough to produce K-ANNS recall close to unity during the construction process (0.95 is enough for the most use-cases). And just like in [26], this parameter can possibly

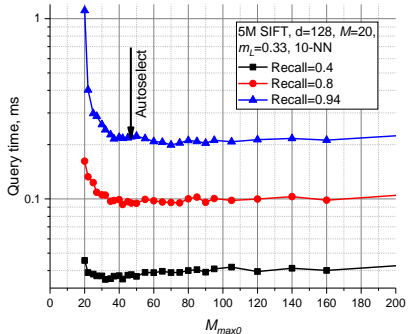

Fig. 6. Plots for query time vs $M_{max0}$ parameter for 5M SIFT learn dataset. The autoselected value $2 \cdot M$ for $M_{max0}$ is shown by an arrow.

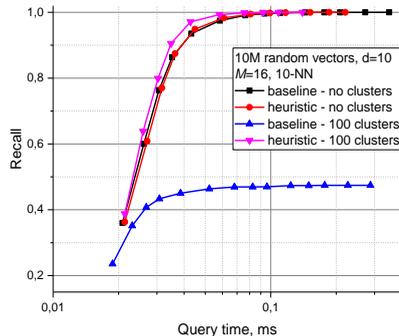

Fig. 7. Effect of the method of neighbor selections (baseline corresponds to alg. 3, heuristic to alg. 4) on clustered (100 random isolated clusters) and non-clustered d=10 random vector data.

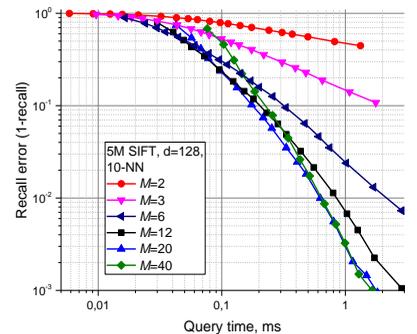

Fig. 8. Plots for recall error vs query time for different parameters of $M$ for Hierarchical NSW on 5M SIFT learn dataset.

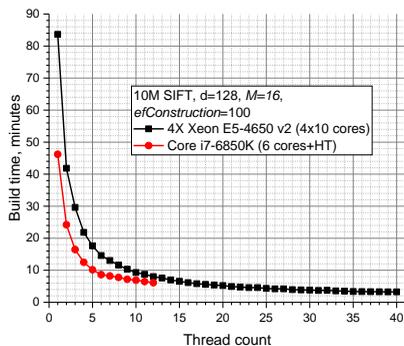
Fig. 9. Construction time for Hierarchical NSW on 10M SIFT dataset for different numbers of threads on two CPUs.

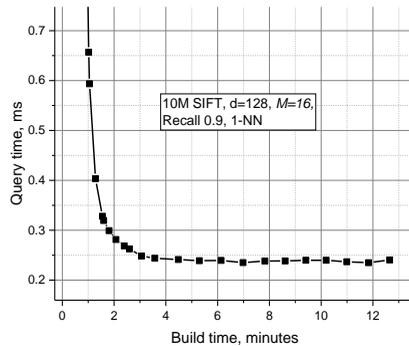
Fig. 10. Plots of the query time vs construction time tradeoff for Hierarchical NSW on 10M SIFT dataset.

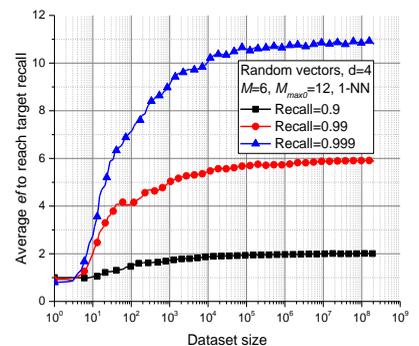
Fig. 11. Plots of the *ef* parameter required to get fixed accuracies vs the dataset size for d=4 random vector data.

be auto-configured by using sample data.

The construction process can be easily and efficiently parallelized with only few synchronization points (as demonstrated in Fig. 9) and no measurable effect on index quality. Construction speed/index quality tradeoff is controlled via the *efConstruction* parameter. The tradeoff between the search time and the index construction time is presented in Fig. 10 for a 10M SIFT dataset and shows that a reasonable quality index can be constructed for *efConstruction*=100 on a 4X 2.4 GHz 10-core Xeon E5-4650 v2 CPU server in just 3 minutes. Further increase of the *efConstruction* leads to little extra performance but in exchange of significantly longer construction time.

## 4.2 Complexity analysis

### 4.2.1 Search complexity

The complexity scaling of a single search can be strictly analyzed under the assumption that we build exact Delaunay graphs instead of the approximate ones. Suppose we have found the closest element on some layer (this is guaranteed by having the Delaunay graph) and then descended to the next layer. One can show that the average number of steps before we find the closest element in the layer is bounded by a constant.

Indeed, the layers are not correlated with the spatial positions of the data elements and, thus, when we traverse the graph there is a fixed probability $p=\exp(-m_L)$ that the next node belongs to the upper layer. However, the search on the layer always terminates before it reaches the element which belongs to the higher layer (otherwise the search on the upper layer would have stopped on a different element), so the probability of not reaching the target on s-th step is bounded by $\exp(-s \cdot m_L)$. Thus the expected number of steps in a layer is bounded by a sum of geometric progression $S=1/(1-\exp(-m_L))$, which is independent of the dataset size.

If we assume that the average degree of a node in the Delaunay graph is capped by a constant $C$ in the limit of the large dataset (this is the case for random Euclid data [48], but can be in principle violated in exotic spaces), then the overall average number of distance evaluations in a layer is bounded by a constant $C \cdot S$, independently of the dataset size.

And since the expectation of the maximum layer index by the construction scales as $O(\log(N))$, the overall complexity scaling is $O(\log(N))$, in agreement with the simulations on low dimensional datasets.

The inital assumption of having the exact Delaunay graph violates in Hierarchical NSW due to usage of approximate edge selection heuristic with a fixed number of neighbors per element. Thus, to avoid stucking into a local minimum the greedy search algorithm employs a backtracking procedure on the zero layer. Simulations show that at least for low dimensional data (Fig. 11, d=4) the dependence of the required *ef* parameter (which determines the complexity via the minimal number of hops during the backtracking) to get a fixed recall saturates with the rise of the dataset size. The backtracking complexity is an additive term in respect to the final complexity, thus, as follows from the empirical data, inaccuracies of the Delaunay graph approximation do not alter the scaling.

Such empirical investigation of the Delaunay graph approximation resilience requires having the average number of Delaunay graph edges independent of the dataset to evidence how well the edges are approximated with a constant number of connections in Hierarchical NSW. However, the average degree of Delaunay graph scales exponentially with the dimensionality [39]), thus for high dimensional data (e.g. d=128) the aforementioned condition requires having extremely large datasets, making such empricial investigation unfeasible. Further analitical evidence is required to confirm whether the resilience of Delaunay graph aproximations generalizes to higher dimensional spaces.

### 4.2.2 Construction complexity

The construction is done by iterative insertions of all elements, while the insertion of an element is merely a sequence of K-ANN-searches at different layers with a subsequent use of heuristic (which has fixed complexity at fixed *efConstruction*). The average number of layers for an element to be added in is a constant that depends on $m_L$:

$$E[l+1] = E[-\ln(unif(0,1)) \cdot m_L] + 1 = m_L + 1 \quad (1)$$

Thus, the insertion complexiy scaling is the same as the one for the search, meaning that at least for relatively low dimensional datasets the construction time scales as $O(N \cdot \log(N))$.

### 4.2.3 Memory cost

The memory consumption of the Hierarchical NSW is mostly defined by the storage of graph connections. The number of connections per element is $M_{max0}$ for the zero layer and $M_{max}$ for all other layers. Thus, the average memory consumption per element is $(M_{max0}+m_L \cdot M_{max}) \cdot bytes\_per\_link$. If we limit the maximum total number of elements by approximately four billions, we can use four-byte unsigned integers to store the connections. Tests suggest that typical close to optimal $M$ values usually lie in a range between 6 and 48. This means that the typical memory requirements for the index (excluding the size of the data) are about 60-450 bytes per object, which is in a good agreement with the simulations.

## 5 PERFORMANCE EVALUATION

The Hierarchical NSW algorithm was implemented in C++ on top of the Non Metric Space Library (nmslib) [49][1], which already had a functional NSW implementation (under name "sw-graph"). Due to several limitations posed by the library, to achieve a better performance, the Hierarchical NSW implementation uses custom distance functions together with C-style memory management, which avoids unnecessary implicit addressing and allows efficient hardware and software prefetching during the graph traversal.

Comparing the performance of K-ANNS algorithms is a nontrivial task since the state-of-the-art is constantly changing as new algorithms and implementations are emerging. In this work we concentrated on comparison with the best algorithms in Euclid spaces that have open source implementations. An implementation of the Hierarchical NSW algorithm presented in this paper is also distributed as a part of the open source nmslib library[1] together with an external C++ memory-efficient header-only version with support for incremental index construction[2].

The comparison section consists of four parts: comparison to the baseline NSW (5.1), comparison to the state-of-the-art algorithms in Euclid spaces (5.2), rerun of the subset of tests [34] in general metric spaces in which NSW failed (5.3) and comparison to state-of-the-art PQ-algorithms on a large 200M SIFT dataset (5.4).

### 5.1 Comparison with baseline NSW

For the baseline NSW algorithm implementation, we used the "sw-graph" from nmslib 1.1 (which is slightly updated compared to the implementation tested in [33, 34]) to demonstrate the improvements in speed and algorithmic complexity (measured by the number of distance computations).

Fig. 12(a) presents a comparison of Hierarchical NSW to the basic NSW algorithm for d=4 random hypercube data made on a Core i5 2400 CPU (10-NN search). Hierarchical NSW uses much less distance computations during a search on the dataset, especially at high recalls.

The scalings of the algorithms on a d=8 random hypercube dataset for a 10-NN search with a fixed recall of 0.95 are presented in Fig. 12(b). It clearly demostrates that Hierarchical NSW has a complexity scaling for this setting not worse than logarithmic and outperforms NSW at any dataset size. The performance advantage in absolute time (Fig. 12(c)) is even higher due to improved algorithm implementaion.

### 5.2 Comparison in Euclid spaces

The main part of the comparison was carried out on vector datasets with use of the popular K-ANNS benchmark ann-benchmark[3] as a testbed. The testing system utilizes python bindings of the algorithms – it consequentially runs the K-ANN search for one thousand queries (randomly extracted from the initial dataset) with preset algorithm parameters producing an output containing recall and average time of a single search. The considered algorithms are:

1. Baseline NSW algorithm from nmslib 1.1 ("sw-graph").
2. FLANN 1.8.4 [6]. A popular library[4] containing several algorithms, built-in in OpenCV[5]. We used the available auto-tuning procedure with several reruns to infer the best parameters.
3. Annoy[6], 02.02.2016 build. A popular algorithm

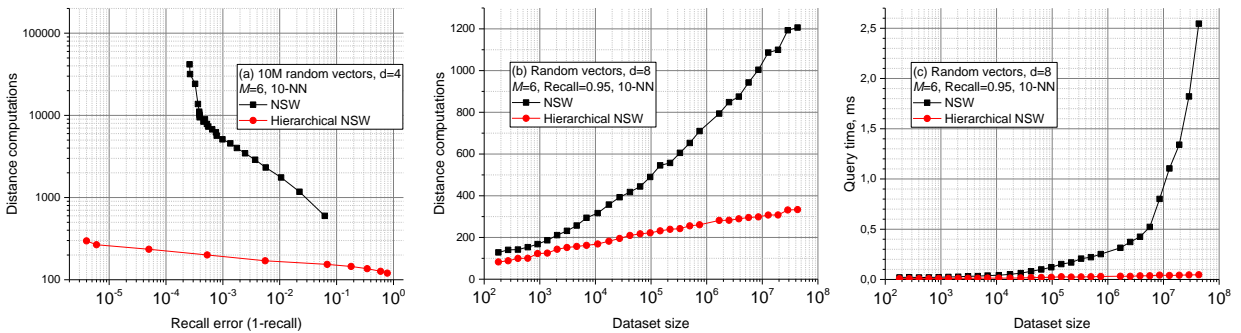

Fig. 12. Comparison between NSW and Hierarchical NSW: (a) distance calculation number vs accuracy tradeoff for a 10 million 4-dimensional random vectors dataset; (b-c) performance scaling in terms of number of distance calculations (b) and raw query(c) time on a 8-dimensional random vectors dataset.

---

[1] https://github.com/searchivarius/nmslib
[2] https://github.com/nmslib/hnsw
[3] https://github.com/erikbern/ann-benchmarks
[4] https://github.com/mariusmuja/flann
[5] https://github.com/opencv/opencv
[6] https://github.com/spotify/annoy

TABLE 1
Parameters of the used datasets on vector spaces benchmark.

| Dataset | Description | Size | d | BF time | Space |
|---|---|---|---|---|---|
| SIFT | Image feature vectors [13] | 1M | 128 | 94 ms | $L_2$ |
| GloVe | Word embeddings trained on tweets [52] | 1.2M | 100 | 95 ms | cosine |
| CoPhIR | MPEG-7 features extracted from the images [53] | 2M | 272 | 370 ms | $L_2$ |
| Random vectors | Random vectors in hypercube | 30M | 4 | 590 ms | $L_2$ |
| DEEP | One million subset of the billion deep image features dataset [14] | 1M | 96 | 60 ms | $L_2$ |
| MNIST | Handwritten digit images [54] | 60k | 784 | 22 ms | $L_2$ |

TABLE 2.
Used datasets for repetition of the Non-Metric data tests subset.

| Dataset | Description | Size | d | BF time | Distance |
|---|---|---|---|---|---|
| Wiki-sparse | TF-IDF (term frequency–inverse document frequency) vectors (created via GENSIM [58]) | 4M | $10^5$ | 5.9 s | Sparse cosine |
| Wiki-8 | Topic histograms created from sparse TF-IDF vectors of the wiki-sparse dataset (created via GENSIM [58]) | 2M | 8 | - | Jensen–Shannon (JS) divergence |
| Wiki-128 | Topic histograms created from sparse TF-IDF vectors of the wiki-sparse dataset (created via GENSIM [58]) | 2M | 128 | 1.17 s | Jensen–Shannon (JS) divergence |
| ImageNet | Signatures extracted from LSVRC-2014 with SQFD (signature quadratic form) distance [59] | 1M | 272 | 18.3 s | SQFD |
| DNA | DNA (deoxyribonucleic acid) dataset sampled from the Human Genome 5 [34]. | 1M | - | 2.4 s | Levenshtein |

4. VP-tree. A general metric space algorithm with metric pruning [50] implemented as a part of nmslib 1.1.
5. FALCONN[7], version 1.2. A new efficient LSH algorithm for cosine similarity data [51].

The comparison was done on a 4X Xeon E5-4650 v2 Debian OS system with 128 Gb of RAM. For every algorithm we carefully chose the best results at every recall range to evaluate the best possible performance (with initial values from the testbed defaults). All tests were done in a single thread regime. Hierarchical NSW was compiled using the GCC 5.3 with -Ofast optimization flag.

The parameters and description of the used datasets are outlined in Table 1. For all of the datasets except GloVe we used the $L_2$ distance. For GloVe we used the cosine similarity which is equivalent to $L_2$ after vector normalization. The brute-force (BF) time is measured by the nmslib library.

Results for the vector data are presented in Fig. 13. For SIFT, GloVE, DEEP and CoPhIR datasets Hierarchical NSW clearly outperforms the rivals by a large margin. For low dimensional data (d=4) Hierarchical NSW is slightly faster at high recall compared to the Annoy while strongly outperforms the other algorithms.

## 5.3 Comparison in general spaces

A recent comparison of algorithms [34] in general spaces (i.e. non-symmetric or with violation of triangle inequality) showed that the baseline NSW algorithm has severe problems on low dimensional datasets. To test the performance of the Hierarchical NSW algorithm we have repeated a subset of tests from [34] on which NSW performed poorly or suboptimal. For that purpose we used a built-in nmslib testing system which had scripts to run tests from [34]. The evaluated algorithms included the VP-tree, permutation techniques (NAPP and bruteforce filtering) [49, 55-57], the basic NSW algorithm and NNDescent-produced proximity graphs [29] (both in pair with the NSW graph search algorithm). As in the original tests, for every dataset the test includes the results of either NSW or NNDescent, depending on which structure performed better. No custom distance functions or special

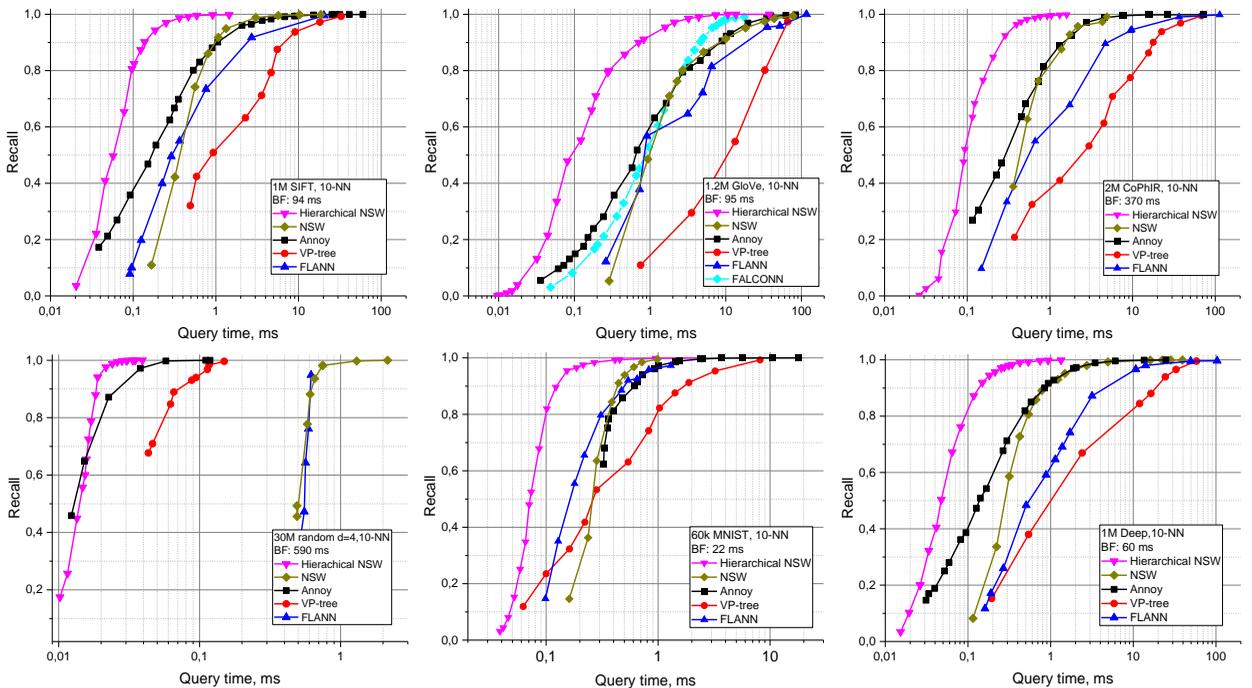

Fig. 13. Results of the comparison of Hierarchical NSW with open source implementations of K-ANNS algorithms on five datasets for 10-NN searches. The time of a brute-force search is denoted as the BF.

[7] https://github.com/FALCONN-LIB/FALCONN

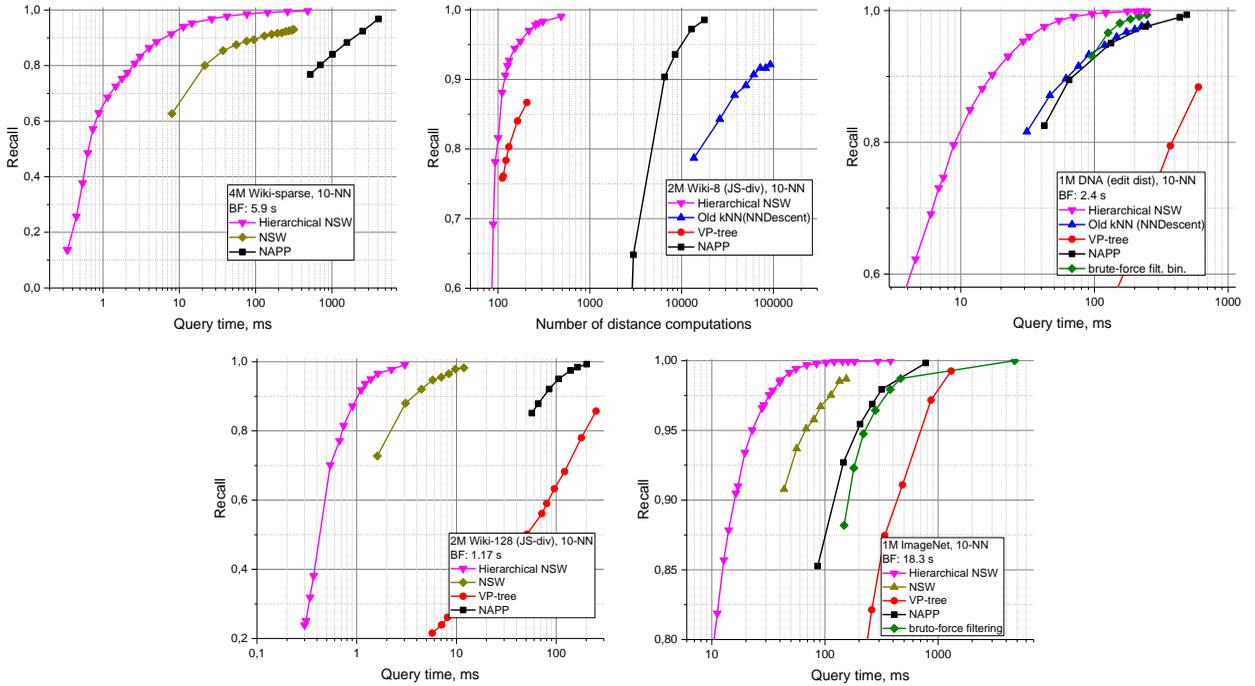

Fig. 14. Results of the comparison of Hierarchical NSW with general space K-ANNS algorithms from the Non Metric Space Library on five datasets for 10-NN searches. The time of a brute-force search is denoted as the BF.

TABLE 3.
Parameters for comparison between Hierarchical NSW and Faiss on a 200M subset of 1B SIFT dataset.

| Algorithm | Build time | Peak memory (runtime) | Parameters |
|---|---|---|---|
| Hierarchical NSW | 5.6 hours | 64 Gb | $M=16$, $efConstruction=500$ (1) |
| Hierarchical NSW | 42 minutes | 64 Gb | $M=16$, $efConstruction=40$ (2) |
| Faiss | 12 hours | 30 Gb | $OPQ64, IMI2x14, PQ64$ (1) |
| Faiss | 11 hours | 23.5 Gb | $OPQ32, IMI2x14, PQ32$ (2) |

memory management were used in this case for Hierarchical NSW leading to some performance loss.

The datasets are summarized in Table 2. Further details of the datasets, spaces and algorithm parameter selection can be found in the original work [34]. The brute-force (BF) time is measured by the nmslib library.

The results are presented in Fig. 14. Hierarchical NSW significantly improves the performance of NSW and is a leader for any of the tested datasets. The strongest enhancement over NSW, almost by 3 orders of magnitude is observed for the dataset with the lowest dimensionality, the wiki-8 with JS-divergence. This is an important result that demonstrates the robustness of Hierarchical NSW, as for the original NSW this dataset was a stumbling block. Note that for the wiki-8 to nullify the effect of implementation results are presented for the distance computations number instead of the CPU time.

### 5.4 Comparison with product quantization based algorithms.

Product quantization K-ANNS algorithms [10-17] are considered as the state-of-the-art on billion scale datasets since they can efficiently compress stored data, allowing modest RAM usage while achieving millisecond search times on modern CPUs.

To compare the performance of Hierarchical NSW

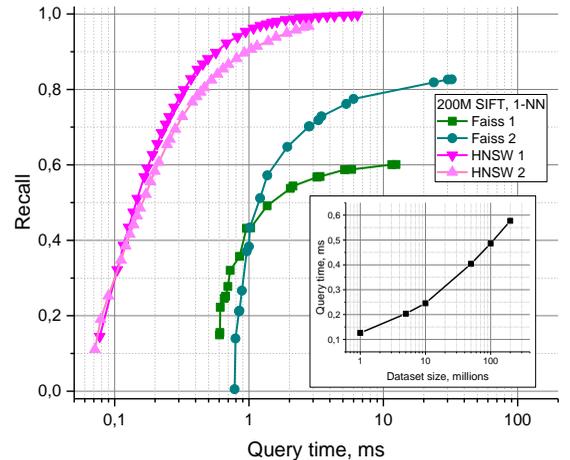

Fig. 15 Results of comparison with Faiss library on the 200M SIFT dataset from [13]. The inset shows the scaling of the query time vs the dataset size for Hierarchical NSW.

against PQ algorithms we used the facebook Faiss library[8] as the baseline (a new library with state-of-the-art PQ algorithms [12, 15] implementations, released after the current manuscript was submitted) compiled with the OpenBLAS backend. The tests where done for a 200M subset of 1B SIFT dataset [13] on a 4X Xeon E5-4650 v2 server with 128Gb of RAM. The ann-benchmark testbed was not feasible for these experiments because of its reliance on 32-bit floating point format (requiring more than 100 Gb just to store the data). To get the results for Faiss PQ algorithms we have utilized built-in scripts with the parameters from Faiss wiki[9]. For the Hierarchical NSW algorithm we used a special build outside of the nmslib with a small memory footprint, simple non-vectorized

---

[8] https://github.com/facebookresearch/faiss 2017 May build. From 2018 Faiss library has its own implementation of Hierarchical NSW.
[9] https://github.com/facebookresearch/faiss/wiki/Indexing-1G-vectors

integer distance functions and support for incremental index construction[10].

The results are presented in Fig. 15 with summarization of the parameters in Table 3. The peak memory consumption was measured by using linux "time –v" tool in separate test runs after index construction for both of the algorithms. Even though Hierarchical NSW requires significantly more RAM, it can achieve much higher accuracy, while offering a massive advance in search speed and much faster index construction.

The inset in Fig. 15 presents the scaling of the query time vs the dataset size for Hierarchical NSW. Note that the scaling deviates from the pure logarithm, possibly due to relatively high dimensionality of the dataset.

## 6 DISCUSSION

By using structure decomposition of navigable small world graphs together with the smart neighbor selection heuristic the proposed Hierarchical NSW approach overcomes several important problems of the basic NSW structure advancing the state-of–the-art in K-ANN search. Hierarchical NSW offers an excellent performance and is a clear leader on a large variety of the datasets, surpassing the opensource rivals by a large margin in case of high dimensional data. Even for the datasets where the previous algorithm (NSW) has lost by orders of magnitude, Hierarchical NSW was able to come first. Hierarchical NSW supports continuous incremental indexing and can also be used as an efficient method for getting approximations of the k-NN and relative neighborhood graphs, which are byproducts of the index construction.

Robustness of the approach is a strong feature which makes it very attractive for practical applications. The algorithm is applicable in generalized metric spaces performing the best on any of the datasets tested in this paper, and thus eliminating the need for complicated selection of the best algorithm for a specific problem. We stress the importance of the algorithm's robustness since the data may have a complex structure with different effective dimensionality across the scales. For instance, a dataset can consist of points lying on a curve that randomly fills a high dimensional cube, thus being high dimensional at large scale and low dimensional at small scale. In order to perform efficient search in such datasets an approximate nearest neighbor algorithm has to work well for both cases of high and low dimensionality.

There are several ways to further increase the efficiency and applicability of the Hierarchical NSW approach. There is still one meaningful parameter left which strongly affects the construction of the index – the number of added connections per layer *M*. Potentially, this parameter can be inferred directly by using different heuristics [4]. It would also be interesting to compare Hierarchical NSW on the full 1B SIFT and 1B DEEP datasets [10-14] and add support for element updates and removal.

One of the apparent shortcomings of the proposed approach compared to the basic NSW is the loss of the possibility of distributed search. The search in the Hierarchical NSW structure always starts from the top layer, thus the structure cannot be made distributed by using the same techniques as described in [26] due to cognestion of the higher layer elements. Simple workarounds can be used to distribute the structure, such as partitioning the data across cluster nodes studied in [6], however in this case, the total parallel throughput of the system does not scale well with the number of computer nodes.

Still, there are other possible known ways to make this particular structure distributed. Hierarchical NSW is ideologically very similar to the well-known one-dimensional exact search probabilistic skip list structure, and thus can use the same techniques to make the structure distributed [45]. Potentially this can lead to even better distributed performance compared to the base NSW due to logarithmic scalability and ideally uniform load on the nodes.

## 7 ACKNOWLEDGEMENTS

We thank Leonid Boytsov for many helpful discussions, assistance with Non-Metric Space Library integration and comments on the manuscript. We thank Seth Hoffert and Azat Davletshin for the suggestions on the manuscript and the algorithm and fellows who contributed to the algorithm on the github repository. We also thank Valery Kalyagin for support of this work.

The reported study was funded by RFBR, according to the research project No. 16-31-60104 mol_a_dk.

---

[10] https://github.com/nmslib/hnsw

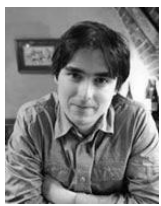
**Yury A. Malkov** received a Master's degree in physics from Nizhny Novgorod State University in 2009, and a PhD degree in laser physics from the Institute of Applied Physics RAS in 2015. He is author of 20+ papers on physics and computer science. Yury currently occupies a position of a Project Leader in Samsung AI Center in Moscow. His current research interests include deep learning, scalable similarity search, biological and artificial neural networks.

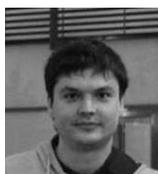
**Dmitry A. Yashunin** received a Master's degree in physics from Nizhny Novgorod State University in 2009, and a PhD degree in laser physics from the Institute of Applied Physics RAS in 2015. From 2008 to 2012 he was working in Mera Networks. He is author of 10+ papers on physics. Dmitry currently woks at Intelli-Vision in the position of a leading research engineer. His current research interests include scalable similarity search, computer vision and deep learning.